\documentclass[usenatbib,useAMS,usegraphicx]{mn2e}

\usepackage{times}
\usepackage[latin1,utf8]{inputenc}
\usepackage{graphicx}
\usepackage[psamsfonts]{amssymb}
\usepackage{xspace}
\usepackage{dcolumn}
\usepackage{bm}
\usepackage{physics}

\newcommand{\VOISE}{\renewcommand{\VOISE}{VOISE\xspace}VOronoi Image
SEgmentation (VOISE)\xspace}
\newcommand{\VR}{\renewcommand{\VR}{VR\xspace}Voronoi region (VR)\xspace}
\newcommand{\VD}{\renewcommand{\VD}{VD\xspace}Voronoi diagram (VD)\xspace}
\newcommand{\CVD}{\renewcommand{\CVD}{CVD\xspace}centroidal Voronoi diagram (CVD)\xspace}
\newcommand{\DT}{\renewcommand{\DT}{DT\xspace}Delaunay triangulation (DT)\xspace}
\newcommand{\HST}{\renewcommand{\HST}{HST\xspace}Hubble Space Telescope (HST)\xspace}
\newcommand{\STIS}{\renewcommand{\STIS}{STIS\xspace}Space Telescope imaging spectrograph (STIS)\xspace}
\newcommand{\ACS}{\renewcommand{\ACS}{ACS\xspace}Advanced Camera for Surveys (ACS)\xspace}
\newcommand{\SBC}{\renewcommand{\SBC}{SBC\xspace}Solar-Blind Channel (SBC)\xspace}
\newcommand{\WFPC}{\renewcommand{\WFPC}{WFPC2\xspace}Wide-Field Planetary
Camera~2 (WFPC2)\xspace}
\newcommand{\UV}{\renewcommand{\UV}{UV\xspace}ultraviolet (UV)\xspace}
\newcommand{\FUV}{\renewcommand{\FUV}{UV\xspace}far-ultraviolet (FUV)\xspace}
\newcommand{\CML}{\renewcommand{\CML}{CML\xspace}Central Meridian Longitude (CML)\xspace}

\newcommand{\vr}{\ensuremath{\mathop{\mathcal{R}}\nolimits}}
\newcommand{\vd}{\ensuremath{\mathop{\mathcal{V}}\nolimits}}
\newcommand{\pt}[1]{\vec{#1}}
\newcommand{\prob}{\ensuremath{\mathop{\mathrm{Prob}}\nolimits}}

\newcommand{\image}{\ensuremath{\mathop{\mathcal{I}}\nolimits}}
\newcommand{\tiled}{\ensuremath{\mathop{\mathcal{T}}\nolimits}}

\renewcommand{\arcsec}{\ensuremath{\mathrm{arc\,sec}}}

\newcommand{\comment}[1]{}

\newlength{\figwidth}
\figwidth=\columnwidth 

\title[VOISE: Automatic Segmentation of Astronomical Images]{The VOISE
Algorithm: a Versatile Tool for Automatic Segmentation of Astronomical Images}
\author[P. Guio and N. Achilleos]{P. Guio\thanks{E-mail:p.guio@ucl.ac.uk}
and N. Achilleos\\ Physics and Astronomy, University College London,
Gower Place, London, WC1E 6BT, United Kingdom}

\begin{document}

\date{Accepted 2009 June 09. Received 2009 June 08; in original form 2009 April 07}

\pagerange{\pageref{firstpage}--\pageref{lastpage}} \pubyear{2009}

\maketitle

\label{firstpage}

\begin{abstract}
The auroras on Jupiter and Saturn can be studied with a high
sensitivity and resolution by the \HST \UV and \FUV \STIS and \ACS
instruments. 
We present results of automatic detection and segmentation
of Jupiter's auroral emissions as observed by \HST \ACS instrument with \VOISE.
\VOISE is a dynamic algorithm for partitioning the underlying pixel grid of
an image into regions according to
a prescribed homogeneity criterion.  The algorithm consists of an iterative
procedure that dynamically constructs a tessellation of the image plane
based on a Voronoi Diagram, until the intensity of the underlying image
within each region is classified as homogeneous. The computed tessellations
allow the extraction of quantitative information about the auroral features
such as mean intensity, latitudinal and longitudinal extents and length
scales. These outputs thus represent a more automated and objective method
of characterising auroral emissions than manual inspection.
\end{abstract}


\section{Introduction}

Planetary auroral emission is a particularly useful diagnostic tool for
the study of magnetospheric processes. 
In the polar regions of the planet, currents
flow into and out of the upper atmosphere and energetic particles
precipitate down the magnetic field lines, impacting on the molecules in
the atmosphere.  At Jupiter and Saturn, the dominant atmospheric species is
hydrogen, which emits in the \UV when excited by auroral electron impact. 
In order to observe this \UV emission we need to use a platform above
the Earth's absorbing atmosphere, such as the \STIS instrument (until the 2004
Saturn campaign), and its successor the \ACS instrument, both on board the
Earth-orbiting \HST. 
From the study of these auroral emissions we can learn a great deal about
magnetospheric processes at these planets without even leaving Earth orbit
\citep{clarke:2002,grodent:2003a,grodent:2003b}.

\comment{
The energy that drives the Earth's auroras ultimately derives from the solar
wind, which is a much more significant momentum and energy source for the
Earth's magnetosphere than planetary rotation. The processes associated with
the Dungey cycle \citep{dungey:1961} drive strong currents parallel to the
magnetic field and energise plasma such that energetic charged particles
flow into the polar ionosphere and generate the auroras that have been
observed on the ground for thousands of years. The location and brightness
of the Earth's auroras are strongly dependent on the prevailing conditions
in the solar wind, e.g.\ \citet{walker:1995}. During periods of high solar 
activity the Earth's
magnetosphere may be buffeted by clouds of plasma emitted from the Sun, and
under certain conditions these can initiate magnetic storms, during which
large amounts of energy are transferred from the solar wind to the
magnetosphere and ionosphere. The Earth's auroras and the associated
magnetospheric dynamics have been studied extensively for decades from both
the ground and space. While such observations have improved our
understanding of auroral formation, much of the underlying physics is still
not well understood, especially with regard to energy release during
substorms.}

The auroras of the outer planets have been studied relatively little in
comparison to the Earth's, however we have advanced our understanding of the
environment of Jupiter and Saturn.  Jupiter's magnetosphere is the
largest cavity in the solar wind flow, i.e.\ within the heliosphere,
and the dynamics of it are largely dominated
by the fast rotation of the planet (period \unit[\sim10]{hr}). It has been
established that the most significant component of Jupiter's auroras, the main
auroral oval, is driven by planetary rotation.  
The moon Io orbits deep within Jupiter's
magnetosphere (orbital radius \unit[\sim6]{R_J} where Jupiter equatorial radius 
$\unit{R_J}=\unit[71492]{km}$), where it is exposed to large tidal forces
as it is subject to the competing gravities of Jupiter and the other
Galilean moons:
Europa, Ganymede, and Callisto.  These tidal forces heat the interior of Io,
such that it has become extremely volcanic, and liberates from its volcanoes
of the order of one tonne per second of sulphur and oxygen plasma into the
neighbourhood of its orbit.  This plasma feels the fast rotation of
Jupiter's magnetic field and is accelerated to the same rotation rate as the
planet.  This fast rotation causes it to `diffuse' away from Jupiter
due to the centrifugal force, and in doing so it slows down its rotation in
an attempt to conserve angular momentum.  This results
in a vast spinning disc of plasma that co-rotates with Jupiter in the
inner region and more slowly in the outer region. Jupiter's magnetic field,
however, attempts to keep all this plasma rotating at the same angular
velocity. This enforcement is mediated by an electric current system, one
component of which results in electrons bombarding
Jupiter's upper atmosphere and causing the main oval auroras
\citep{hill:1979,vasyliunas:1983}. 

Jupiter also exhibits two other classes of auroral
feature: the moon footprints and the polar auroras. The moon footprints are
spots of auroral emission that are magnetically linked to the Galilean
moons.  The brightest spot is that associated with Io, and it also exhibits
a tail, or `wake' that traces eastward (in the direction of rotation) 
around the planet.  The exact details
of the processes that lead to the formation of the footprints are still not
understood, but it is thought that they are a result of the moons'
interaction with the rotating magnetic field and the acceleration of newly
created plasma to the speed of the rotating field via Alfvén waves
\citep{gurnett:1981b,bonfond:2007}. 
The polar auroras are
still somewhat of an enigma, as they are the most highly variable component
of Jupiter's auroras in terms of brightness.  The location and variation of
these auroral emissions make it possible that at least some of them may be
caused by the solar wind interaction, although this has yet to be proved
conclusively. Interestingly, recent work \citep{pallier:2001} shows good
evidence that Jupiter's polar cap boundary, i.e.\ the perimeter of the region
where planetary field lines open into the solar wind, is coincident at times
with some of these polar emissions.

\comment{
Saturn's auroras are different again from those of the Earth and Jupiter.
Like the Earth's they exhibit major changes in their morphology in response
to changes in the solar wind \citep{cowley:2005} but they are also influenced
by the rapid
rotation of the planet (\unit[\sim11]{hr}). Saturn exhibits an auroral oval
that occasionally takes the form of a spiral and sometimes appears to be
constructed from several separate `arcs'.  In one striking set of images the
auroras expanded to fill almost the entire dawn side of the polar cap in
response to a sharp increase in the dynamic pressure of the solar wind
observed by Cassini-Huygens as the spacecraft approached the planet in 2004.
One of the goals
of the 2007 \HST observation campaign was to determine whether Saturn's
auroras are driven by internal or external processes, or a combination 
\citep{clarke:2002}.}

The preceding discussions highlight an important link between the
geometrical properties and locations of auroral features, and the physical
processes of their origin. Previous  observational studies, e.g.\
\citet{pallier:2001}, have used correlation techniques (spatial filtering)
to emphasise auroral
features at selected length scales. In this paper, we present a new automated
technique for producing maps of the entire range of length scales and other
properties of features in a given auroral image.

Segmentation refers to the process of partitioning a digital image into
multiple regions (sets of pixels). The goal of segmentation is to simplify
and/or change the representation of an image into something that is more
physically meaningful and easier to analyse. Image segmentation is typically
used to locate objects and boundaries (lines, curves, etc.) in images, and
different approaches have been developed especially for the purpose of 
medical imaging \citep{pham:2000}. 
In the astrophysical context different methods have also been used in image
processing and analysis. Amongst others, we mention the watershed
transform \citep{platen:2007}, the  adaptive spatial binning
\citep{cappellari:2003} and Bayesian image reconstruction
\citep{cabrera:2008} .

In this paper we present the \VOISE algorithm, a method for segmentation
of an image into regions based on the partitioning of the support of the
image (the set of data values over all pixels) into a set of disjoint convex
polygons.
Each polygon is a \VR and the boundary points of the collection of
the \VR{s} is a \VD. The \VD is sometimes also known as
a Dirichlet tessellation and the \VR{s} are also called Dirichlet regions, 
Meijering cells, Thiessen polytopes, or Voronoi polygons \citep{okabe:2000}. 
The set of \VR{s} is determined
by a finite number of points (one point inside each \VR) called
\emph{seeds}, sometimes called \emph{germs} or \emph{generators}.
\VOISE is an iterative and self-organising algorithm
\citep{jantsch:1980,nicolis:1989} for
automatic segmentation of an image, based on adaptive construction of a \VD.
It consists of dynamically subdividing and merging a network of convex polygons
according to the information contained in the underlying image.

In section~\ref{sec:vd}, we give a detailed introduction about the \VD, and 
in section~\ref{sec:voise}, the \VOISE algorithm is developed.
The behaviour of auroras is frequently studied through manual inspection of
two-dimensional image data. In section~\ref{sec:application}, we apply
the \VOISE algorithm to images of Jupiter's auroras from the \HST \ACS
instrument with the \SBC. In section~\ref{sec:discussion}, we discuss
the results of the segmentation.  We expect the \VOISE algorithm to 
provide more objective and more efficient analyses than visual
inspection, especially when such analysis relies on statistics of large
volumes of data, whose manual processing may be prone to human error.

Additional natural byproducts of the \VOISE algorithm include
compression of data and noise-limited detection of features. These aspects
will be explored in a future study. It is the purpose of the present paper
to introduce the fundamentals of the technique, together with its application
in auroral physics study.

\section{Voronoi Diagram}
\label{sec:vd}

The \VD is one of a few truly interdisciplinary concepts with relevant
material to be found in fields as different as anthropology, 
astronomy, 
ecology, 
physics, 
and urban and regional planning \citep{okabe:2000}.
Let us begin by considering a set of seeds
$S=\{\pt{s}_1,\pt{s}_2,\cdots,\pt{s}_n\}$,
where $\pt{s}_i$
are distinct points in the plane $\mathbb{R}^2$.
The \VR associated with seed $\pt{s}_i$, noted $\vr(\pt{s}_i)$, is the set
of points $\pt{p}$ in the plane $\mathbb{R}^2$ that are nearer to $\pt{s}_i$
than to any other seeds $\pt{s}_j$, with respect to the Euclidean distance
$d$, formally
\begin{align}
\vr(\pt{s}_i)=\{\pt{p}\in\mathbb{R}^2\,|\,d(\pt{p},
                 \pt{s}_i)<d(\pt{p},\pt{s}_j), 
\quad\pt{s}_j\in S\backslash\pt{s}_i\},
\label{eq:vr}
\end{align}
and its closure, noted $\overline{\vr}(\pt{s}_i)$, is defined as
\begin{align}
\overline{\vr}(\pt{s}_i)=\{\pt{p}\in\mathbb{R}^2\,|\,d(\pt{p},
                 \pt{s}_i)\leq d(\pt{p},\pt{s}_j), 
\quad\pt{s}_j\in S\backslash\pt{s}_i\},
\label{eq:vrclosed}
\end{align}
The \VD of a set of seeds $S$, noted $\vd(S)$, is the graph or `skeleton' 
formed by the polygonal boundaries of the \VR{s} of all seeds in $S$, i.e.\
\begin{align}
\vd(S) = \bigcap\limits_{\pt{s}_i\in S}\overline{\vr}(\pt{s}_i).
\label{eq:vd}
\end{align}
Since the intersection of any \VR{s} is empty and the union of the closure
of all \VR{s} is the plane, the \VD is called a tessellation of the plane.

In the terminology of mathematical morphology, the \VR $\vr(\pt{s}_i)$ is
called the \emph{influence zone} or the \emph{dominance region} of $\pt{s}_i$,
and $\vd(S)$, the \emph{skeleton by influence zones}. Note that the definitions
given in Eqs.~(\ref{eq:vr}--\ref{eq:vd}) are still valid for any space where a
distance has been defined .
For instance, in the three-dimensional Euclidean space, the \VR are convex
polyhedra and the \VD consists of the skeleton formed by the 
boundaries of the \VR{s}.

\begin{figure}
\centering
\includegraphics[width=0.99\figwidth]{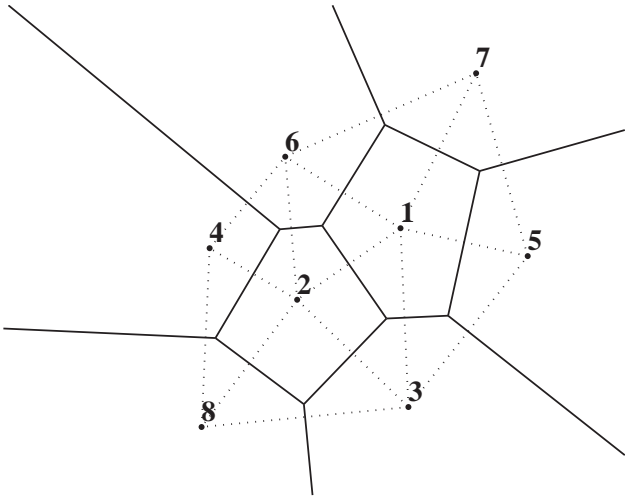}
\caption{Illustration of a planar \VD \emph{(solid lines)}  and its dual \DT
\emph{(dotted lines)} of a set $S$ consisting of eight seeds (labelled 1 to
8).} 
\label{VD}
\end{figure}

Figure~\ref{VD} illustrates the \VD for a set of eight seeds in the plane.
Two seeds are said to be neighbours if their associated \VR{s} share a common
edge.

In addition to its geometrical representation, the \VD can also be
identified by an abstract representation called a Voronoi \emph{graph}.
The Voronoi graph is a \emph{planar} graph in two-dimensional space.
The graph $G(S)=(V,E)$ represents the topological structure
of the neighbouring \VR{s}. The nodes (or vertices) $V$ and lines
(or edges) $E$ of the  graph $G$ represent respectively
the seeds and the unordered pairs of nodes that represent links connecting
two seeds. 
In Figure~\ref{VD}, the neighbours of seed $\pt{s}_1$ consists of the set
$\mathcal{N}(\pt{s}_1)=\{\pt{s}_2,\pt{s}_3,\pt{s}_5,\pt{s}_7,\pt{s}_6\}$.
Thus we can deduce that the pairs $\{\pt{s}_1,\pt{s}_2\},\{\pt{s}_1,\pt{s}_3\},
\{\pt{s}_1,\pt{s}_5\},\{\pt{s}_1,\pt{s}_7\}$ and
$\{\pt{s}_1,\pt{s}_6\}$ are links of the graph associated to the node
$\pt{s}_1$. 

The data structure consisting of the
neighbour list for all seeds $\mathcal{N}(S)=\{\mathcal{N}(\pt{s}_1),
\mathcal{N}(\pt{s}_2),\ldots,\mathcal{N}(\pt{s}_n)\}$ thus represents
the Voronoi graph.

The set $\mathcal{N}(S)$ also characterises the \DT, 
which is obtained by connecting all the pairs of neighbour seeds.
The nodes and faces (triangles) of the \DT correspond respectively to the
polygons and nodes of the \VD. The \DT is called the \emph{dual} graph
representation of the primal graph of the \VD. 

The \DT are widely used in various applications such as finite element
methods and interpolation schemes, owing to the fact that, once the set of
seeds $S$ has been defined, the \DT optimises the geometric `compactness' of
the triangulation. In particular, the \DT maximises the minimum angles of all
its element triangles. Also the \DT is the unique triangulation where 
the circumscribed circle of each triangle does not contain
any other vertices.

Note also that a \VR may be either a bounded or an unbounded region. When a
\VR is bounded the number of edges and vertices are equal, when unbounded
there is one vertex less than there are edges. In Figure~\ref{VD}, the
regions $\vr(\pt{s}_1)$ and $\vr(\pt{s}_2)$ are bounded while all the
others are unbounded. The polygon $\mathcal{P}$ consisting of the vertices 
$\{\pt{s}_4,\pt{s}_8,\pt{s}_3,\pt{s}_5,\pt{s}_7,\pt{s}_6\}$,
which are the seeds of all the unbounded regions, is called the convex hull of the set $S$.

The construction of the \VD and the \DT is a fundamental problem in
computational geometry, and many approaches have been proposed to compute
them, such as the incremental method, the divide-and-conquer method and
the plane sweep method \citep{okabe:2000}.

An image in this context is a real-valued function $\image(\pt{p})$ defined
for $\pt{p}$, pixels in the image plane: a two-dimensional lattice space
$\Omega\subset\mathbb{N}^2$. 
In the rest of this paper, we restrict the set of seeds $S$ to
be distinct points in the image plane $\Omega$.
And accordingly, the \VR and \VD are defined as their restriction to the image
plane $\Omega$.

\section{\VOISE Algorithm}
\label{sec:voise}

\VOISE is an iterative and self-organising algorithm, and
consists of up to four different phases explained hereafter.

\subsection{Initialisation Phase}
\label{init}
A small number of seeds are initially \emph{randomly} drawn according to 
the uniform probability distribution over the image plane $\Omega$, and
the corresponding \VD is constructed.

The generation of points in space according to a probability distribution
is called a stochastic point process. For a uniform probability
distribution over a bounded region the process is the binomial point
process, where for any subset of the region, the probability
distribution for the number of points within the subset is given by
the binomial distribution.
Asymptotically, if we let the bounded region tend toward the infinite region
$\mathbb{R}^2$ while keeping the density of points constant, the process tends
toward the homogeneous Poisson point process and the tessellation is called
a Poisson-Voronoi tessellation \citep{okabe:2000}.

The set of \VR{s} represents a `tiling' of the image plane $\Omega$.
Let us consider an operator $f(\image,\pt{s}_i)$ which associates, to
each seed $\pt{s}_i$, a unique value depending on the region $\vr(\pt{s}_i)$,
and the intensity of the image pixels within that region, i.e.\ the
restriction $\image{|}_{\vr(\pt{s}_i)}$ of $\image$ to $\vr(\pt{s}_i)$,
with values in the set $\{\image(\pt{p}), p\in\vr(\pt{s}_i)\}$.
Then a tiled image $\tiled_f$ can be constructed by `filling' each polygon
with the value given by applying the operator $f(\image,\pt{s}_i)$, 
formally
\begin{align}
\tiled_f(p) = f(\image,\pt{s}_i),\quad\mbox{for}\quad \pt{s}_i\in S
\quad\mbox{and}\quad
p\in\vr(\pt{s}_i). 
\label{eq:op}
\end{align}
In order to calculate a tiled approximation of an image with a set of seeds
$S$, one can for instance use statistical averages such as the arithmetic
mean or the median for the operator $f$.

\subsection{Dividing phase}
\label{divide}
The dividing phase is an iterative process and is the main process for
`evolving' the \VD. First we need to define a measure
of \emph{homogeneity} for the set of pixels of the underlying image within
each \VR of the diagram. We define the merit function $\chi$ of the image
$\image$ in the region associated to seed $\pt{s}_i$ as
\begin{align}
\chi\left(\image,\pt{s}_i\right) &= 
\frac{\max\limits_{\pt{p}\in\vr(\pt{s}_i)}\left[\image(\pt{p})\right]-
      \min\limits_{\pt{p}\in\vr(\pt{s}_i)}\left[\image(\pt{p})\right]}{
\|\chi(\image,S)\|}\label{eq:chi}\\
\|\chi(\image,S)\| &= \max\limits_{\pt{s}_i\in S}
\left[\max\limits_{\pt{p}\in\vr(\pt{s}_i)}\left[\image(\pt{p})\right]-
      \min\limits_{\pt{p}\in\vr(\pt{s}_i)}\left[\image(\pt{p})\right]\right]
			\label{eq:normchi}
\end{align}

The numerator of the right hand side in Eq.~(\ref{eq:chi}) is a
`max-min' measure of the variation of the image intensity within the
region $\vr(\pt{s}_i)$. The numerator defined in Eq.~(\ref{eq:normchi}) 
is the $L$-infinity norm of this max-min measure of the image intensity
for the entire \VD.
The values of the measure of homogeneity $\chi\left(\image,\pt{s}_i\right)$
are, by definition, bounded in the interval $[0,1]$.
The minimum value of $\chi$ corresponds to the most homogeneous 
\VR and conversely, the maximum value is for the least homogeneous
region. Note that the merit function $\chi$ defined in 
Eqs.~(\ref{eq:chi}--\ref{eq:normchi}) is only one of many possibilities. Any
alternative merit function can easily be plugged in \VOISE.

\begin{figure}
\centering
\includegraphics[width=0.99\figwidth]{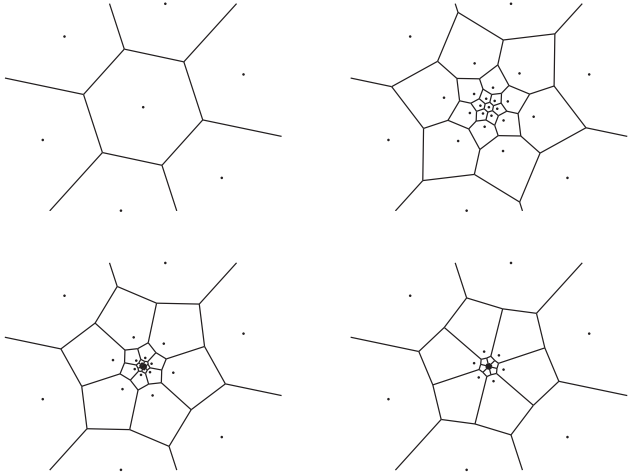}
\caption{Illustration of the method to add seeds for different weightings.
Seeds are added to the central hexagonal region (\emph{upper left} panel) 
recursively for three iterations, adding 18~seeds. The weights are
$w_s=1/4,w_v=3/4$ (\emph{upper right} panel), $w_s=1/2,w_v=1/2$ 
(\emph{lower left} panel) and $w_s=3/4,w_v=1/4$ (\emph{lower left} panel).} 
\label{addSeeds2VR}
\end{figure}

When a cell is not homogeneous according to the chosen homogeneity threshold
$\chi_m$, i.e.\ 
\begin{align}
\chi(\image,\pt{s}_i)\geq\chi_m,
\end{align}
seeds are added within this region. 
The number of seeds added
corresponds to the number of vertices in the polygon. Each added seed is the
barycentre of the seed of the \VR with weight $w_s$, and
a vertex of the polygon with weight $w_v$, corresponding to a fixed
relative distance from the seed.
The most probable number of vertices/edges of a typical cell of a
Poisson-Voronoi diagram is six \citep{okabe:2000}, in other words
the most probable polygon is an hexagon.
Figure~\ref{addSeeds2VR} illustrates the method for three iterations
starting with the hexagon in the centre of the upper left panel and
iteratively over the innermost hexagon, adding six~seeds at each iteration, and
for three weightings, $(w_s=1/4,w_v=3/4)$ in the upper right panel,  
$(w_s=1/2,w_v=1/2)$ in the lower let panel and $(w_s=3/4,w_v=1/4)$ in the lower
right panel. In the last case where the seeds are added nearest the seed of the
original region, the newly created central region is very small compared to the
newly created surrounding regions. In \VOISE we set these parameters to
$w_s=5/8$ and $w_v=3/8$ which provides a good compromise between quick
convergence and comparable size of new regions.

This iterative and deterministic algorithm of 
adding new seeds ensures the regular growth of regions such that 
self-similar structure across different length scales is emphasised
(fractals with recursive definition such as the Koch flake are classic
examples of self-similarity).

The threshold $\chi_m$ is a \emph{dynamic} and \emph{self-organising} parameter
estimated for each new iteration of a dividing phase using quantile
statistics on the sample $\chi_i$ of the measure of homogeneity, i.e.\
\begin{align}
\chi_i=\{\chi\left(\image,\pt{s}_i\right),\pt{s}_i\in S\}.
\end{align}
$\chi_m$ is defined as the value of the homogeneity measure 
for which a fraction $p_\mathrm{D}/100$ of the polygons are homogeneous,
i.e.\ such that $\chi(\image,\pt{s}_i)<\chi_m$.
Thus defining the probability \prob, that a given polygon has its
value $\chi(\image,\pt{s}_i)$ below the threshold $\chi_m$, we can write
the following
\begin{align}
\prob\left(\chi(\image,\pt{s}_i)<\chi_m\right) = \frac{p_\mathrm{D}}{100}
\end{align}
Simultaneously, the probability that a given polygon has its
value $\chi(\image,\pt{s}_i)$ exceeding the threshold $\chi_m$ is given by
\begin{align}
\prob\left(\chi(\image,\pt{s}_i)\geq\chi_m\right) = 1-\frac{p_\mathrm{D}}{100}
\end{align}
The value $\chi_m$ is obtained at each iteration as the inverse of the
cumulative histogram of the sample $\chi_i$ for the value $p_\mathrm{D}$
(see middle panel in Figure~\ref{param}).
Since the cumulative histogram is a monotonic increasing function, smaller
values for $p_\mathrm{D}$ give smaller value for the threshold $\chi_m$
and conversely.

Note that as the subdividing process takes place, the norm 
$\|\chi(\image,S)\|$ is becoming smaller, and the measure
of homogeneity is becoming more stringent. The set of homogeneity
measures will accumulate towards the superior value $\sup(\chi)=1$.
The dividing phase is iterated until all \VR{s} are classified as
homogeneous, \emph{or} the size of the polygon does not allow the further
addition of new seeds. This is characterised by enforcing a minimum distance
$d_m$ between two seeds.

For a large value $p_\mathrm{D}$ of the percentile (\unit[70\mbox{--}80]{\%}),
the threshold $\chi_m$ increases, and thus the dividing scheme will only apply
to the \unit[(100-p_\mathrm{D})]{\%} most inhomogeneous regions according
to the merit function $\chi$. 
With the merit function defined as in Eq.~(\ref{eq:chi}), a
large value of $p_\mathrm{D}$ will only 
allow the dividing process to lock onto the most contrasted regions and
will not necessarily be able to detect low-contrasted or noisy features
of the image.  Conversely, for a small value $p_\mathrm{D}$ of the percentile 
(\unit[60\mbox{--}70]{\%}), the dividing process will be more systematic and
uniform over the entire image but will not account for the most contrasted 
features of the image.
However, it is important to emphasise that the \VOISE algorithm is not
limited to the present merit function and any other merit function that
would help characterise a feature, e.g.\ texture, noise properties, can
be plugged in.

The minimum distance $d_m$ between seeds is the parameter which
limits the length scale of the features one wishes to detect. This parameter
should be chosen according to the size of the smallest resolvable 
features.

\subsection{Merging phase}
\label{merge}
At this stage, any \VR is either homogeneous with respect to the
prescribed threshold $\chi_m$, \emph{or} its shape does not allow to
add any new seed as described in the previous section and fulfil the minimum
distance requirement $d_m$ between neighbouring seeds.

The merging phase is also an iterative process. It consists of removing
unnecessary `fine' regions where adjacent \VR have very similar
characteristics, e.g.\ the averaged intensity $\mu_i$ of the image for the
region associated to seed $\pt{s}_i$ as calculated by Eq.~(\ref{eq:op}) with a
specified operator $f$, in the following the median of the set of values. 

A seed $\pt{s}_i$ is defined as redundant and ought to be removed when the
following conditions are fulfilled:
\begin{itemize}
\item[(i)] The region $\vr(\pt{s}_i)$ is homogeneous with respect to
a chosen homogeneity threshold $\chi_m$, i.e.\
\begin{align}
\chi(\image,\pt{s}_i)<\chi_m
\end{align}
The value for the threshold is estimated
at each iteration from the cumulative histogram of the merit function sample
$\chi_i$ from a prescribed $p_\mathrm{M}$-th percentile, in the same way as 
for $p_\mathrm{D}$ in section~\ref{divide}
\begin{align}
\prob\left(\chi(\image,\pt{s}_i)<\chi_m\right) = \frac{p_\mathrm{M}}{100}
\end{align}
\item[(ii)] amongst its neighbouring seeds $\pt{s}_j\in\mathcal{N}(\pt{s}_i)$
that are embedded in homogeneous region, the relative difference between the 
averaged values $\mu_i$ and $\mu_j$ is less than a prescribed threshold
$\Delta\mu$, formally 
\begin{align}
|\mu_i-\mu_j|<\Delta\mu|\mu_i|.
\end{align}
Two such regions are said to be identical with respect to the averaged
intensity $\mu$.
\item[(iii)] the total length $\mathcal{L}$ of the edges shared with non
homogeneous neighbours normalised to the perimeter $\mathcal{P}$ of the
region does not exceed a prescribed threshold $\Delta\mathcal{H}$, formally
\begin{align}
\mathcal{L}/\mathcal{P}<\Delta\mathcal{H}.
\end{align}
\end{itemize}

At each iteration, all seeds are checked for redundancy and removed when
fulfilling the three conditions above. After a set of seeds has been
removed, some regions have grown to absorb the seeds that have been
removed. In such regions, the merit function $\chi$ can only increase (and
may become non homogeneous), and the averaged intensity is changing. 
Thus during the merging process, the three conditions become more difficult
to fulfil for any seeds and the process stops by itself
(`self-regulating').

The percentile of the merging phase, $p_\mathrm{M}$,  does not need to be
identical to the percentile of the dividing phase, $p_\mathrm{D}$,
but should be larger in order to relax the homogeneity criteria and
conversely smaller in order to strengthen the homogeneity criteria (see
Figure~\ref{param}).

The merging phase does not disturb the overall representation of the image but
provides a better approximation with fewer seeds. When a seed is removed,
the neighbouring polygons reorganise themselves to fill the region
associated to the removed seed. We noted that the polygons tend to organise
according to the geometrical shapes in the image. 
The final organisation is made independent of the initialisation
phase by using the merging phase.

Our tests on the image presented in the next section show that 
different initial distributions of seeds do not
significantly change the final \VD. Typically the numbers of polygons, i.e.\
the number of seeds in the final \VD is affected by less than
\unit[4\mbox{--}5]{\%}. 
The typical difference in area of the \VR{s} for different realisations
is less than \unit[10\mbox{--}30]{pixels^2}. And the typical difference in
intensity is less that \unit[5\mbox{--}10]{\%}. Here we computed
the difference in area of the \VR{s} between two realisations as follows. 
In a given realisation $R_1$ of the \VOISE algorithm, each polygon
$\vr(\pt{s}_i^1)$ with seed $\pt{s}_i^1$ has an area $\mathcal{A}_i^1$. 
To compute the difference in polygon area between $R_1$ and a second
realisation $R_2$, we determine for each seed $\pt{s}_i^1$ from $R_1$, the
polygon $\vr(\pt{s}_j^2)$ from $R_2$, with seed $\pt{s}_j^2$ and area
$\mathcal{A}_j^2$ which contains
this seed, i.e.\ such that $\pt{s}_i^1\in\vr(\pt{s}_j^2)$. 
The difference in area is then $\left|\mathcal{A}_i^1-\mathcal{A}_j^2\right|$.

\subsection{Regularisation phase}
\label{regularise}
The tessellation obtained from the divide-and-merge phase can be
`regularised' through an iterative relaxation process.
This is achieved by computing iteratively a new \VD where each seed $\pt{s}_i$
is replaced by the mass centroid (centre-of-mass) $\pt{\xi}_i$ of the
polygon $\vr(\pt{s}_i)$ with a density function $\rho$
\begin{align}
\pt{\xi}_i=\frac{\sum\limits_{\pt{p}\in\vr(\pt{s}_i)}\pt{p}\rho(\pt{p})}
{\sum\limits_{\pt{p}\in\vr(\pt{s}_i)}\rho(\pt{p})} 
\label{eq:com}
\end{align}

When the position of each seed coincides with the centre-of-mass of the
polygonal region, the partitioning is called a \CVD tessellation and 
the method is known as Lloyd's method.
The \CVD corresponds to an optimally regular and uniform tessellation with
a minimum-energy configuration, in the sense that it
minimises the norm of the second-order inertial moment of each polygon.
Asymptotically, the \CVD approaches a uniform hexagonal-like lattice
\citep{du:1999} for constant density function.

The density function $\rho$ has an important role in the regularisation as
it is seen in \citet{du:1999}.
Let us consider the case where the image itself is used as density
function, i.e.\ $\rho(\pt{p})$ is the value of the image data in the pixel
at position $\pt{p}$. After the dividing and merging phases, the polygons
are distributed such that they minimise the merit function $\chi$ defined in 
Eqs.~(\ref{eq:chi}--\ref{eq:normchi}), and the associated density function
should be, to a good approximation, constant within each polygon.

Note that the regularisation process is controlled in two manners. 
The iteration process stops by itself when the relative distance between
the centre-of-mass and the seed of any region, i.e.\
\begin{align}
|\pt{\xi}_i-\pt{s}_i|\quad\mbox{for all}\quad\pt{s}_i\in S,
\end{align}
is less than one pixel. One pixel is enough since the positions of
the centres-of-mass are rounded off to an integer number in order to be
in the image plane $\Omega$. 
In addition, and if required, it is possible to specify a maximum number of
iterations of the relaxation process.

The regularisation scheme could also be used as a preconditioning to the
first iteration(s) of the dividing phase. Following the initialisation
phase where a limited number of seeds are drawn randomly, a regularisation
phase with the image as density function tends to move the seeds to positions
where the image data tends to form local `peaks'.

\subsection{Implementation}
Note that due to the dynamic nature of the algorithm where seeds and added
and removed, it would be quite inefficient to construct the entire \VD
every time the seed configuration is modified, and a more efficient approach
is to use an incremental algorithm which operates only on a given seed and
its neighbours (see Figure~\ref{addSeeds2VR}). In addition the \VR{s} are
needed for the
points in the image plane, which consists of a discrete and well organised
set of points.  We have chosen to use an incremental algorithm for discrete
\VD using a structure similar to the one described in \cite{sequeira:1997},
which is designed for seeds located on the image plane $\Omega$, i.e.\
at discrete positions.

\VOISE requires a set of parameters to be initialised. These
consist of $p_\mathrm{D}$ and $d_m$ for the dividing phase, $p_\mathrm{M}$,
$\Delta\mu$ and $\Delta\mathcal{H}$ for the merging phase, and an 
optional maximum number of iterations for the regularisation phase,
corresponding to partial regularisation. 

\section{Application to \HST image}
\label{sec:application}

\begin{figure}
\centering
\includegraphics[width=1.0\figwidth]{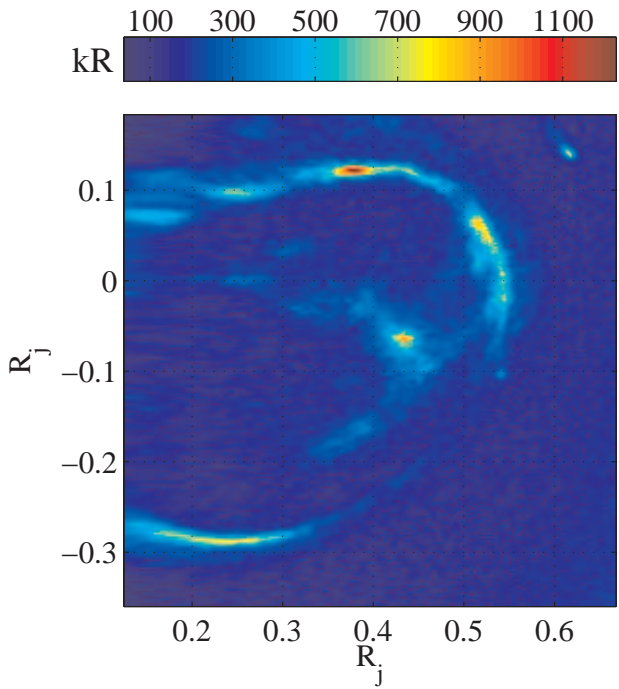}
\caption{Sub-image extracted from the polar projection of Jupiter’s \UV auroral
emission computed from an \HST \ACS/\SBC image taken 
on February 21, 2007 at \unit[16{:}03{:}58]{UT}. The size of the image is
\unit[256\times256]{pixels} and the axis units are expressed in Jupiter's
equatorial radius (\unit[71492]{km}). The origin corresponds to the
projection of the northern rotational pole of Jupiter and the colour-coded
intensity is in \unit{kR}.}
\label{hst}
\end{figure}

Jupiter's \UV auroras have been observed for many years but recently
high-resolution images using the \HST \STIS and \ACS instruments  have
revealed highly dynamic auroral emissions that are controlled to a
significant degree by the solar wind \citep{clarke:2002,nichols:2008}.

Figure~\ref{hst} shows an image of Jupiter's \UV auroral emission in the
Northern hemisphere taken on February 21, 2007 at \unit[16{:}03{:}58]{UT}
and projected onto a northern polar view.

The raw data were collected with the \SBC channel of
the \ACS through the F125LP long-pass filter, with short wavelength
cutoff of \unit[125]{nm}. This filter mostly excludes the \chem{H} 
Lyman-$\alpha$ band. The instrument consists of a \unit[1024\times1024]{} 
MultiAnode Microchannel Array detector with an average scale of
\unit[\sim0.032]{\arcsec\;pixel^{-1}}, such that the overall field of view is
\unit[35\times31]{\arcsec^2}.
During each 100s exposure the blurring introduced by planetary rotation of
any co-rotating features is \unit[\sim1]{deg} at the \CML. The
raw images are reduced: corrected for geometric distortion and scaled 
to \unit[0.0250]{\arcsec\;pixel^{-1}}, flat-fielded and dark-count calibrated
using the latest calibration files available from the Space Telescope Science
Institute. The images are then converted from \unit{counts\;pixel^{-1}} to 
\unit{kR} of \chem{H_2} and Lyman-$\alpha$ emission (where \unit[1]{kR}
represents a photon source flux of \unit[10^9]{cm^{-2}\;s^{-1}} radiating
into \unit[4\pi]{steradians}) using the conversion factor 
$\unit[1]{kR} = \unit[1.473\cdot10^{-3}]{counts\;s^{-1}\;pixel}$
\citep{nichols:2008}.

The image in Figure~\ref{hst} is a reduced polar projection computed
using the meta data provided with the original image and the NASA Navigation
and Ancillary Information Facility SPICE system \citep{acton:1996}, and
assuming the light emission is from an infinitesimally thin shell located
\unit[200]{km} above the \unit[1]{Bar} level. The \unit[1]{Bar} level
corresponds to a oblate spheroid of eccentricity 0.3 and
semi-major axis \unit[71492]{km}. 

The \CML of Jupiter at the time of the observation is \unit[166.77]{deg}. 
The orientation of the image is such that the $x$-axis corresponds to
the \CML meridian, and the dimensions along the axes are in units of Jupiter's
radius ($\mathrm{R_J}=\unit[71492]{km}$).

A non-continuous auroral oval is clearly identified, as well as
several patches of emission within associated to plasma injection and
particle precipitation.
The footprint of Io is seen in the upper right part of the image at
coordinates $(0.6\mathrm{R_J},0.15\mathrm{R_J})$. The footprint of 
the magnetospheric cusp is seen at $(0.4\mathrm{R_J},-0.05\mathrm{R_J})$.

\begin{figure}
\centering
\includegraphics[width=0.99\figwidth]{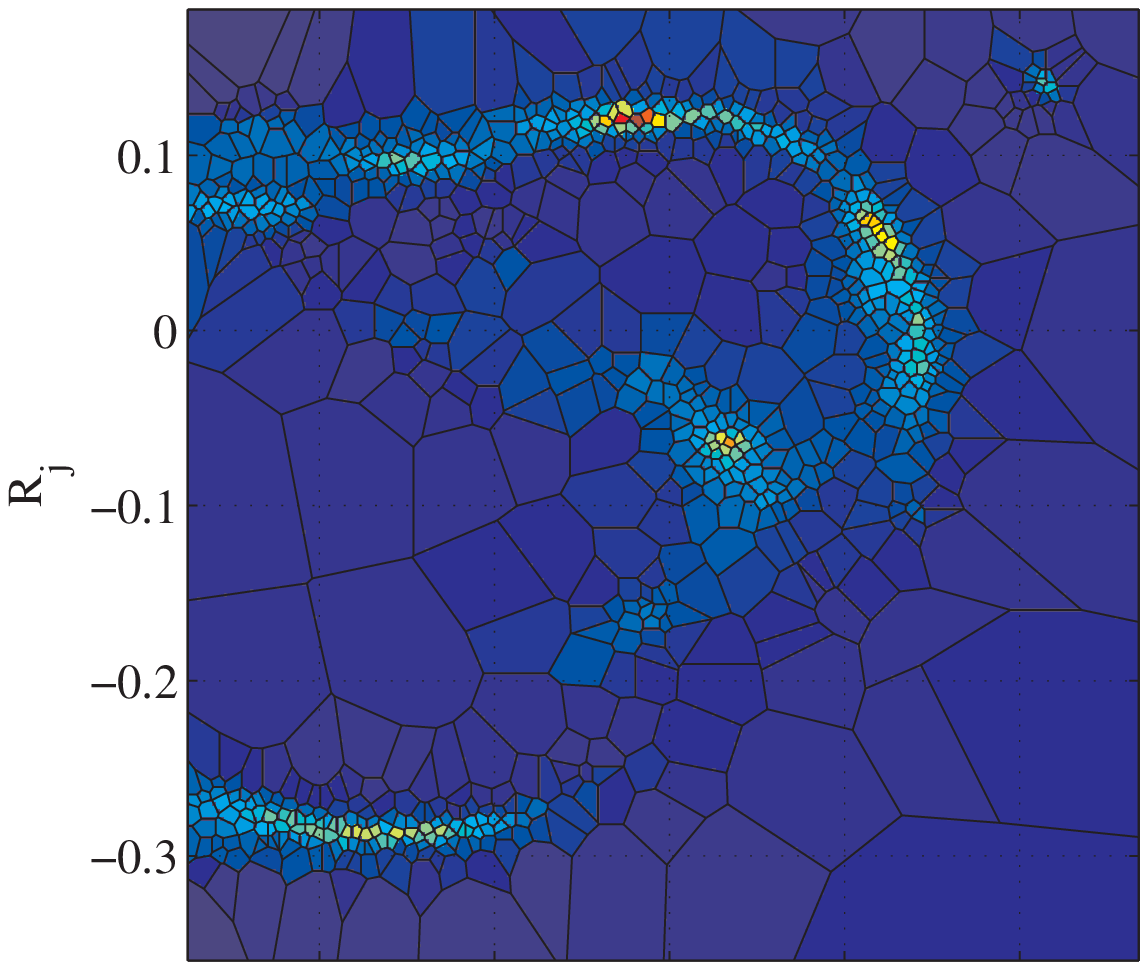}\\[0pt]
\includegraphics[width=0.99\figwidth]{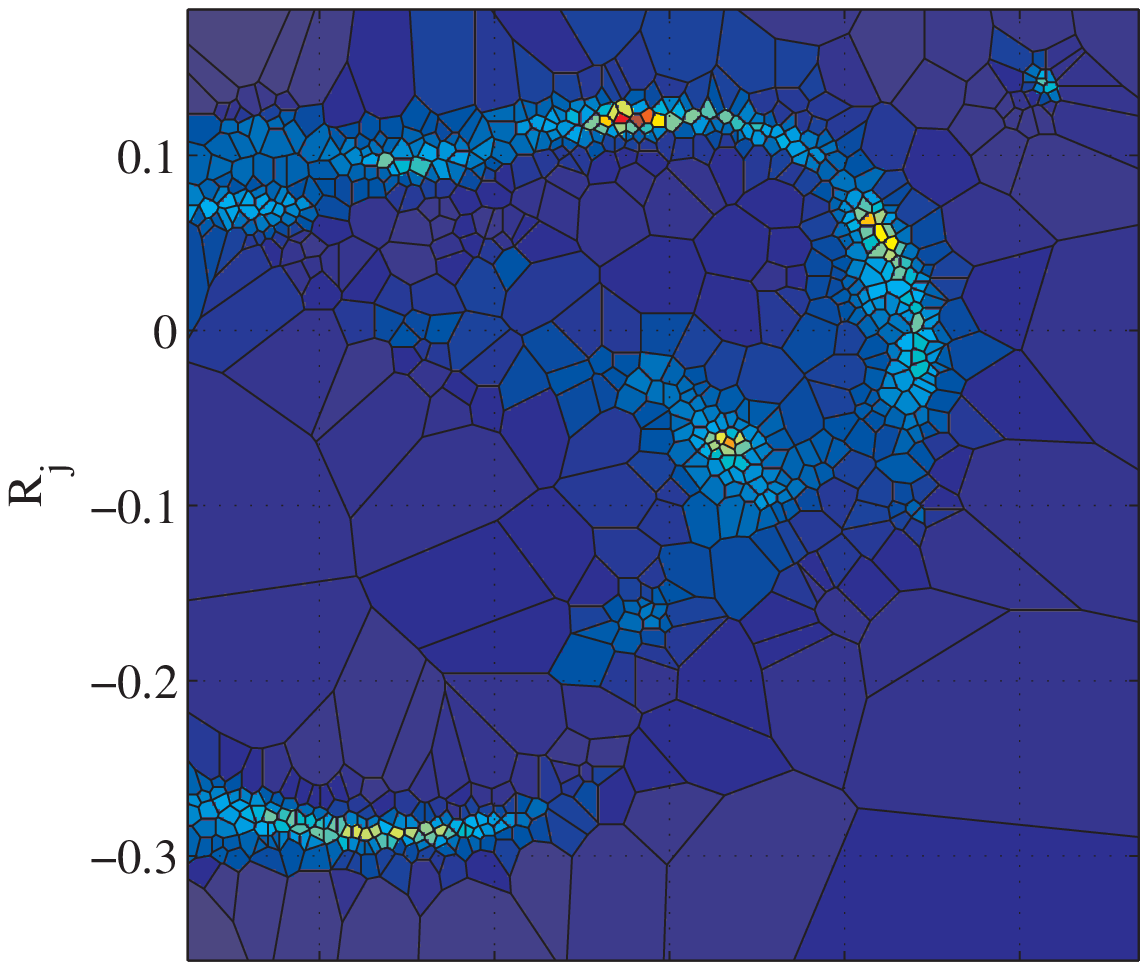}\\[0pt]
\includegraphics[width=0.99\figwidth]{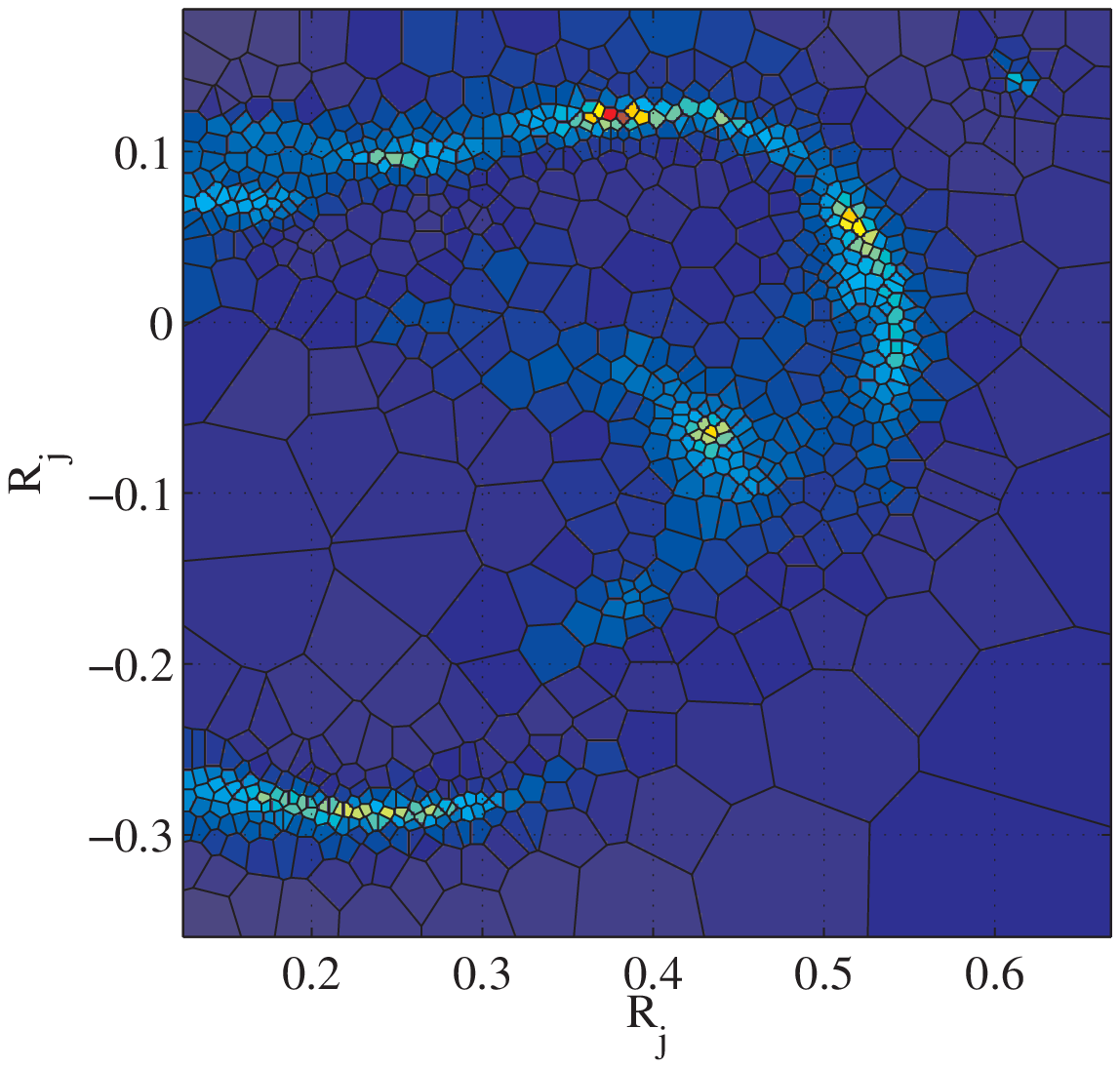}
\caption{Illustration of the \VOISE segmentation algorithm on the image
shown in Figure~\ref{hst}. \emph{Top} panel: result of the dividing phase; 
\emph{middle} panel: result of the merging phase; \emph{bottom} panel: 
result of the regularisation phase.}
\label{algo}
\end{figure}

Figure~\ref{algo} illustrates the result 
for the three phases: dividing, merging and regularisation of
the \VOISE algorithm
applied to the \HST image presented in Figure~\ref{hst}.
The `tiled' images are constructed by filling each polygon with
the median value of the pixels that lie within each polygon.
and the \VD consists of the black solid lines.
The \VD was initialised with twelve random seeds.

\begin{figure}
\centering
\includegraphics[width=0.99\figwidth]{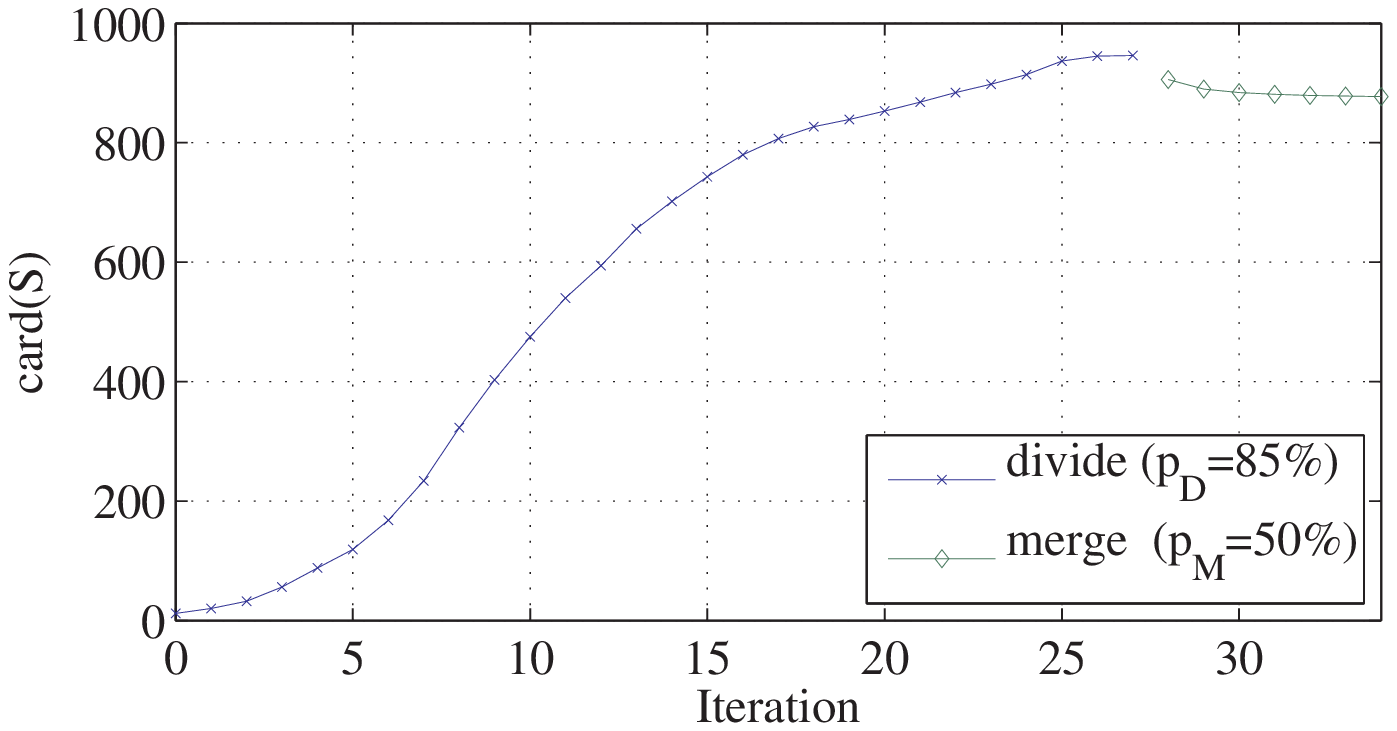}\\[0pt]
\includegraphics[width=0.99\figwidth]{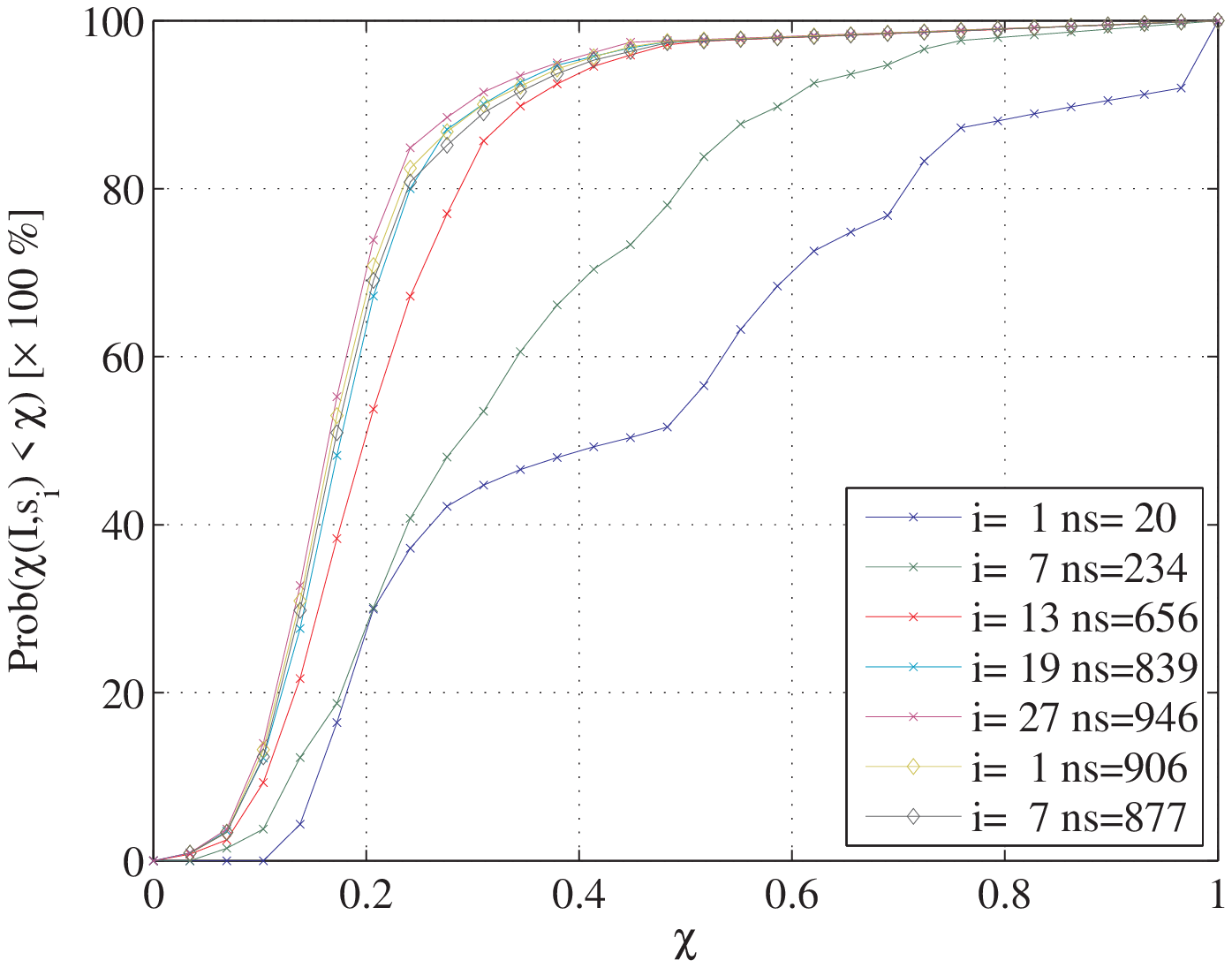}\\[0pt]
\includegraphics[width=0.99\figwidth]{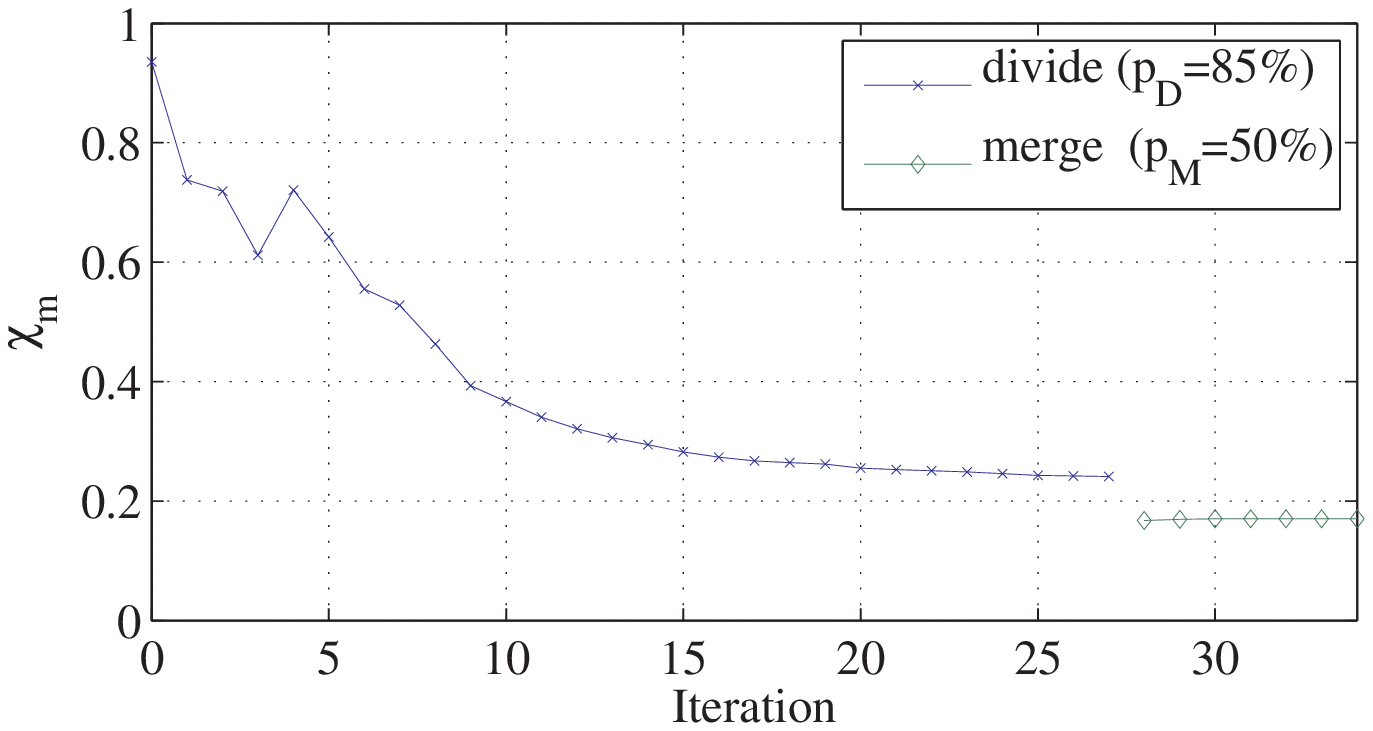}
\caption{\emph{Top} panel: the number of seeds, $\mathrm{card}(S)$, as
function of iteration; \emph{middle} panel: the cumulative histogram of the
sample $\chi_i$ associated to the homogeneity merit function $\chi$, as
defined in Eqs.~(\ref{eq:chi}--\ref{eq:normchi}), for a
selection of iterations of dividing and merging phases.
These iterations are labelled in the legend together with the number of seeds;
\emph{bottom} panel: the threshold $\chi_m$ derived from the cumulative
histogram with the prescribed $p_\mathrm{D}=\unit[85]{\%}$ and
$p_\mathrm{M}=\unit[50]{\%}$
as function of iteration. \emph{Solid} lines with \emph{cross} markers refer
to dividing phase data while \emph{solid} lines with \emph{diamond} markers
refer to merging phase data. }
\label{param}
\end{figure}

The upper image shows the result of the dividing phase. The minimum distance
between two seeds was set to $d_m^2=\unit[7]{pixels^2}$ and the
homogeneity of the \VD by a
requirement of a percentile $p_\mathrm{D}=\unit[85]{\%}$. 
The percentile $p_\mathrm{D}=\unit[85]{\%}$ is directly read from the ordinate
of the cumulative histogram of the sample $\chi_i$ in the middle panel of
Figure~\ref{param}, and the corresponding $\chi_m$ is read from the abscissa.
The legend gives the iteration number and the number of seeds for the
corresponding histogram.
The dividing phase converged with the prescribed parameters
in 27~iterations and about 950~seeds, as seen in the top panel of
Figure~\ref{param}. Initially the cumulative histogram has the signature of
a uniform probability distribution, and as the dividing process takes place 
it converges to a probability distribution where most of the regions
($p_\mathrm{D}=\unit[85]{\%}$) have their homogeneity criteria
$\chi<\chi_m\sim0.25$.

The middle image shows the result of the merging phase. The threshold for
relative similarity is $\Delta\mu=\unit[20]{\%}$, the ratio of non-homogeneous
boundaries is set to $\Delta\mathcal{H}=\unit[30]{\%}$ and the homogeneity of
the \VD is set to a prescribed percentile $p_\mathrm{M}=\unit[50]{\%}$. As for
the dividing phase, the percentile $p_\mathrm{D}$ is directly read in the
ordinate of the cumulative histogram of the merit function $\chi$ in the middle
panel of Figure~\ref{param}, and the corresponding $\chi_m$ is read in the
abscissas.
The merging phase consists of seven iterations after
which the process stops by itself,  as no seeds can fulfil the redundancy
conditions described.

The lower image shows the result of the regularisation phase. The 
tessellation results from two relaxation iterations, i.e.\ replacing 
the seed of each polygon by the centre-of-mass of the associated 
region twice. Our experience shows that one or two iterations are in
general enough, due to the rounding off of the coordinates of the
centre-of-mass, given by Eq.~(\ref{eq:com}), to integer values
corresponding to points of the image plane $\Omega$.

It is worth noting that the \VOISE algorithm based on a statistical
distribution of seeds evolves dynamically according to self-organising of the
information in the image as represented by the changing homogeneity
threshold $\chi_m$.

\section{Discussion} 
\label{sec:discussion}

Note that any quantity or measure could be calculated for each
\VR, based on either the properties of the polygons, 
the statistical properties of the underlying image or a combination of both.

Many analytical properties of the \VR have been determined assuming a
Poisson-Voronoi tessellation. These include
the probability distribution of the perimeter segments, of the angles at
the vertices, the polygon area, the polygon to be $n$-sided, to name a few
\citep{okabe:2000,hilhorst:2008}.
Any deviations of the statistics of a property of the \VD from the
probability distribution of the Poisson-Voronoi tessellation 
is an indication of seed distribution related to some real point pattern, or
clustering as defined in \citet{kaufman:2005}.

Such `tiled' images can thus be used to extract quantitative information
about structures present in the image, but also potentially to clean an
image from unwanted parts such as contamination or defects.

\begin{figure}
\centering
\includegraphics[width=1.0\figwidth]{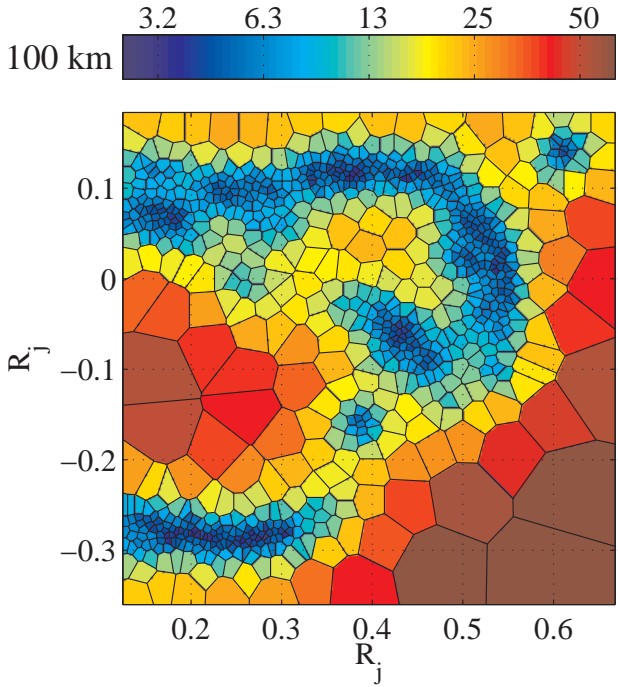}
\caption{Length scale estimated from the geometry of the Voronoi polygons
obtained after regularisation (bottom panel in Figure~\ref{algo}).
The length scale within a \VR is defined as the square root of the area of the
polygon. The length scale is colour-coded in logarithmic scale.}
\label{ls}
\end{figure}

Figure~\ref{ls} shows the length scale 
calculated as the square root of the number of pixels in each Voronoi
polygon times the edge of a pixel in units of \unit[100]{km} at the planet.
This map of length scales allows post-processing such as selection of points
or regions of the image according to a prescribed rule.

\begin{figure}
\centering
\includegraphics[width=0.99\figwidth]{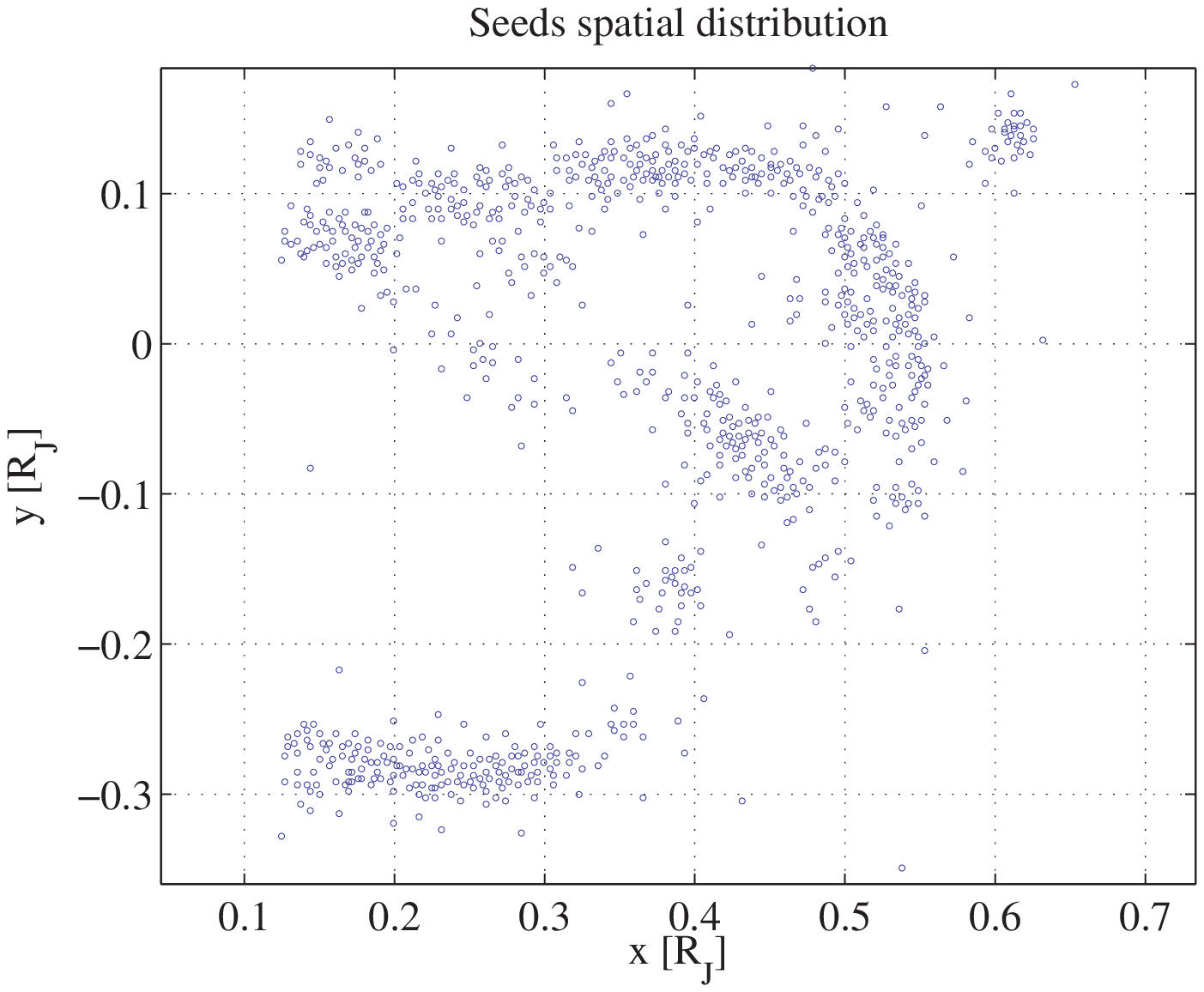}\\[0pt]
\includegraphics[width=0.99\figwidth]{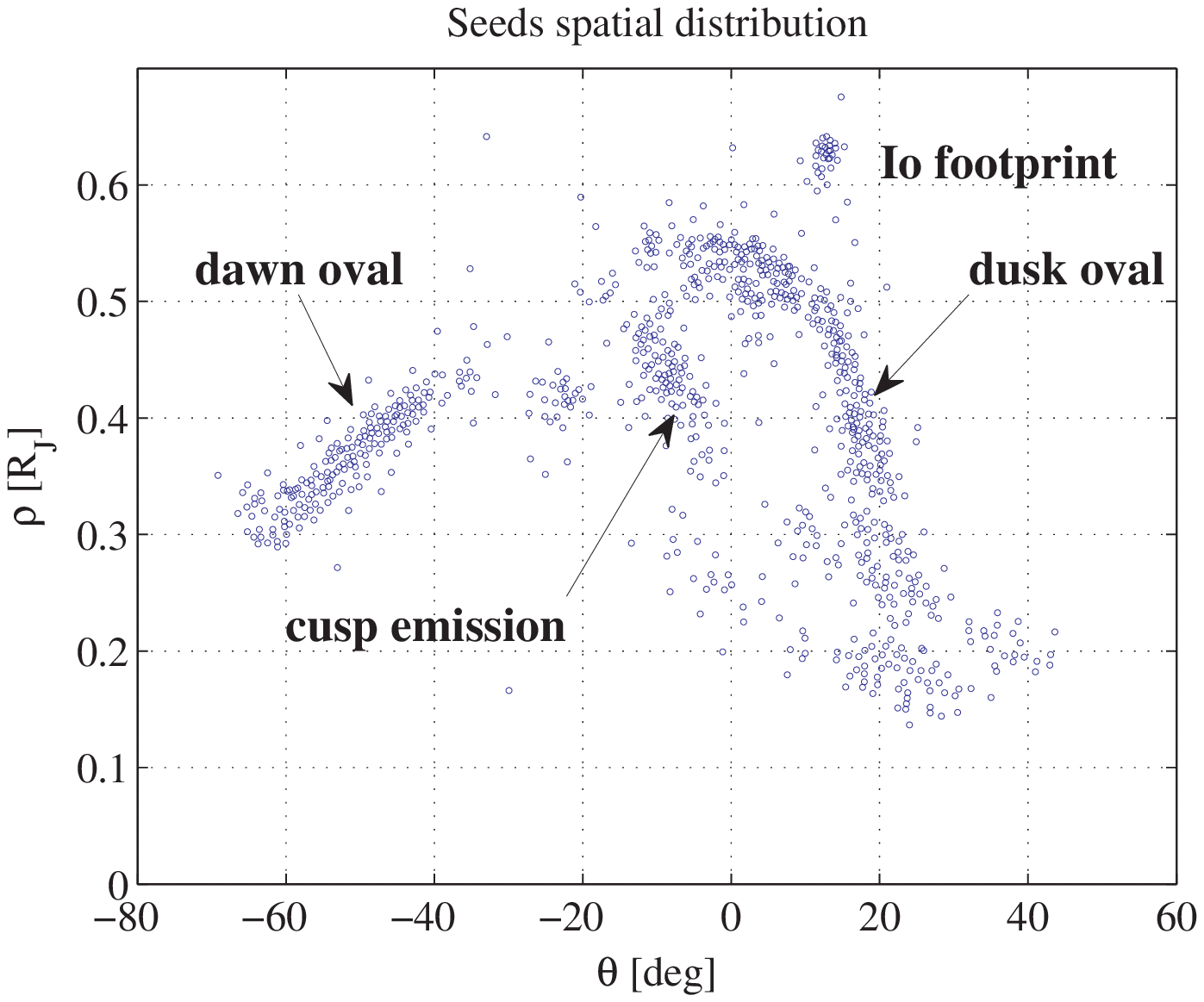}
\caption{Spatial distribution of the seeds for the segmentation obtained 
the merging phase (\emph{middle} panel in Figure~\ref{algo}) expressed 
in Cartesian (\emph{upper} panel) and in polar (\emph{lower} panel)
coordinate systems.}
\label{seeddist}
\end{figure}

Figure~\ref{seeddist} shows the spatial distribution of the seeds of the \VD 
of the middle panel of Figure~\ref{algo} in Cartesian and polar coordinate
systems. 
Let us consider a group of neighbouring seeds, i.e.\ where each
polygon in the group shares at least one edge with another polygon. Such
group of seeds is defined as a cluster if the polygons in the group are
similar with respect to a given property, or combination of properties
\citep{kaufman:2005}.  Such properties could be, for example, purely
geometrical (e.g.\ area, perimeter, length scale), or related to data values
from the underlying image inside polygons (e.g.\ median intensity, mean
intensity). It is important to point out that geometrical properties need
not be used, although they would be a natural choice for detection of
auroral features.
Four such clusters with similar median intensity and length scales are
identifiable and annotated in the polar
coordinate system: (i) the footprint of Io, (ii) the dusk side segment of
the main auroral oval, (iii) the dawn side segment of this oval, and (iv) 
a structure associated to cusp emission. 
The number of polygons per unit area could be considered a crude
representation of the `density of information' for the image.

Our experiments with mapping polar emissions have frequently detected
arc-like auroral features inside the main oval, but running almost parallel
to it. These will be the subject of another study. However they are similar
to the auroral signatures of the polar cap boundary discussed by
\citet{pallier:2001}.

It is interesting to note that the clustering `compactness'
of the seeds is dependent on the coordinate systems. For
instance the footprint of Io in the upper right corner appears more
clustered in polar coordinates.
The scattered and relatively low number of seeds allows
the high compression of information (with loss) achieved by
the \VOISE tessellation.
The original size of the image shown in Figure~\ref{hst} is
\unit[256\times256]{pixels} with pixel intensity coded over
\unit[4]{bytes}, thus requiring \unit[256^2\cdot4]{bytes}. 
The tessellation contains \unit[877]{points}. Coding the median intensity
over \unit[4]{bytes} and each coordinate over \unit[2]{bytes} requires
\unit[877\cdot(4{+}2{+}2)]{bytes}, leading to a
compression factor of 37; alternatively coding the coordinates over
\unit[4]{bytes} reduces the compression factor to 25.
Quantitative estimates of parameters or features such as discontinuities
in the auroral oval, bifurcation of auroral arcs, polar cap position, footprint
of moons, could easily be obtained by fitting the scattered relevant seeds to
an analytical model. Some of these features are indicated in
Figure~\ref{seeddist}.
The automatic detection of the limb of the planet to locate accurately and
objectively the centre of the planet disc is another potential for application
of the \VOISE algorithm.

A common concern that runs through a wide range of disciplines is the
examination of the spatial occurrence of a particular phenomenon, or 
\emph{point patterns}, such as \emph{clustering analysis}
\citep{kaufman:2005}.
The analysis of point patterns of the seeds (or spatial occurrence) can be 
done to divide the \VD into component parts. For instance it was
suggested that individual points may be considered to belong to one of five
types: isolated points, members of a curvilinear structure, members of a
cluster with an empty interior, or members of either the boundary or the
interior of a non-empty cluster \citep{okabe:2000}. The edges of the
polygons here aid this classification.

Another application is multi-temporal or multi-source image analysis and
especially the detection of event defined as a structure modification
between two images. This can be achieved by comparing the \VD of two images
generated by \VOISE according to the following definition.

Let us consider two sets of seeds $S_1$ and $S_2$ and their associated \VD.
Two seeds $\pt{s}_i\in S_1$ and $\pt{s}_j\in S_2$ are \emph{equivalent}
seeds when the seed $\pt{s}_i$ belongs to $\vr(\pt{s}_j)$, and simultaneously
the seed $\pt{s}_j$ belongs to $\vr(\pt{s}_i)$. The regions associated to
seeds $\pt{s}_i$ and $\pt{s}_j$ are
identical or equivalent when the averaged intensities $\mu_i$ and $\mu_j$
of the two images for the region associated to seed $\pt{s}_i$ and $\pt{s}_j$
are similar with respect to an imposed threshold, while an event is detected
between the two images if the averaged intensities are not similar.
The application to analysis of auroral features and their temporal
variability is evident, and will be pursued in future studies.

\section{Conclusion}

We have presented \VOISE, a self-organising dynamic
algorithm for the automatic segmentation of an image based on adaptive
construction of a \VD according to some information contained in the
underlying image.

The \VOISE algorithm has been applied to the detection of \UV auroral
emission regions on Jupiter as observed by the \ACS/\SBC instrument on board
the \HST.

In a following paper, we plan to do a statistical survey of the length
scales, morphologies and variability of the auroras for an extensive set of
\HST \ACS/\SBC images of Jupiter, and look for scale-invariant characteristics,
a property of dynamic systems known as self-organised criticality.  At
present, sequences of typically about \unit[18]{images} (covering time
intervals about or less than \unit[2]{hrs}) are available for 55~different days
during the year~2007, providing many different views of both the northern
and southern \UV auroras.  The geometrical nature of auroral emissions is
known to be linked with their physical origin. Bright concentrated `arcs'
are the result of strong electric currents flowing between the ionosphere
and magnetosphere, while more extended, diffuse emissions arise from
scattering of the magnetospheric particles onto trajectories which intercept
the planetary atmosphere.  Since the macroscopic behaviour of the solar wind
displays spatial and temporal scale-invariance which are characteristic of
self-organised criticality \citep{chapman:1998,chapman:2001}, one would expect
similar macroscopic behaviour of
the  auroral activity to exhibit related spatial and temporal
scale-invariance.

The \VOISE algorithm could easily be extended to three-dimensional data
(discrete volume composed of voxels) to perform automatic volume segmentation.

The applicability of the \VOISE algorithm is not limited to segmentation of
auroral images but can in practice be used for any types of image data with
a suitable homogeneity merit function.

\section*{Acknowledgements}

This work is based on observations with the NASA/ESA Hubble Space
Telescope, obtained at the Space Telescope Science Institute, which is
operated by AURA for NASA.

We thank the \HST auroral campaign team the the Planetary Atmospheres and
Space Science Group at Boston University for providing the data of
Jupiter and for fruitful discussions in the presented method.

We also would like to thank the referee for helpful suggestions and
comments.

\bibliography{abbrevs,research,books}

\bsp

\label{lastpage}

\end{document}